\documentclass[useAMS,usenatbib]{mn2e}\usepackage{ifthen}
\usepackage{graphicx,amsmath,amssymb}
\newif\ifhighlight
\highlightfalse

\usepackage{color}
\usepackage[usenames,dvipsnames]{xcolor}

\ifhighlight
\newcommand{\highlight}{ \color{BrickRed} }
\else
\newcommand{\highlight}{ }
\fi

\newboolean{astroph}
\setboolean{astroph}{false}

\newboolean{mn2e}
\setboolean{mn2e}{true}

\newboolean{aastex}
\setboolean{aastex}{false}

\newcommand*{\dif}{\ensuremath{\mathrm{d}}}

\newcommand{\sinc}{\ensuremath{\mathrm{sinc}}}

\newcommand{\figSamplingtheorem}{
  \begin{figure}
    \includegraphics[width=3in]{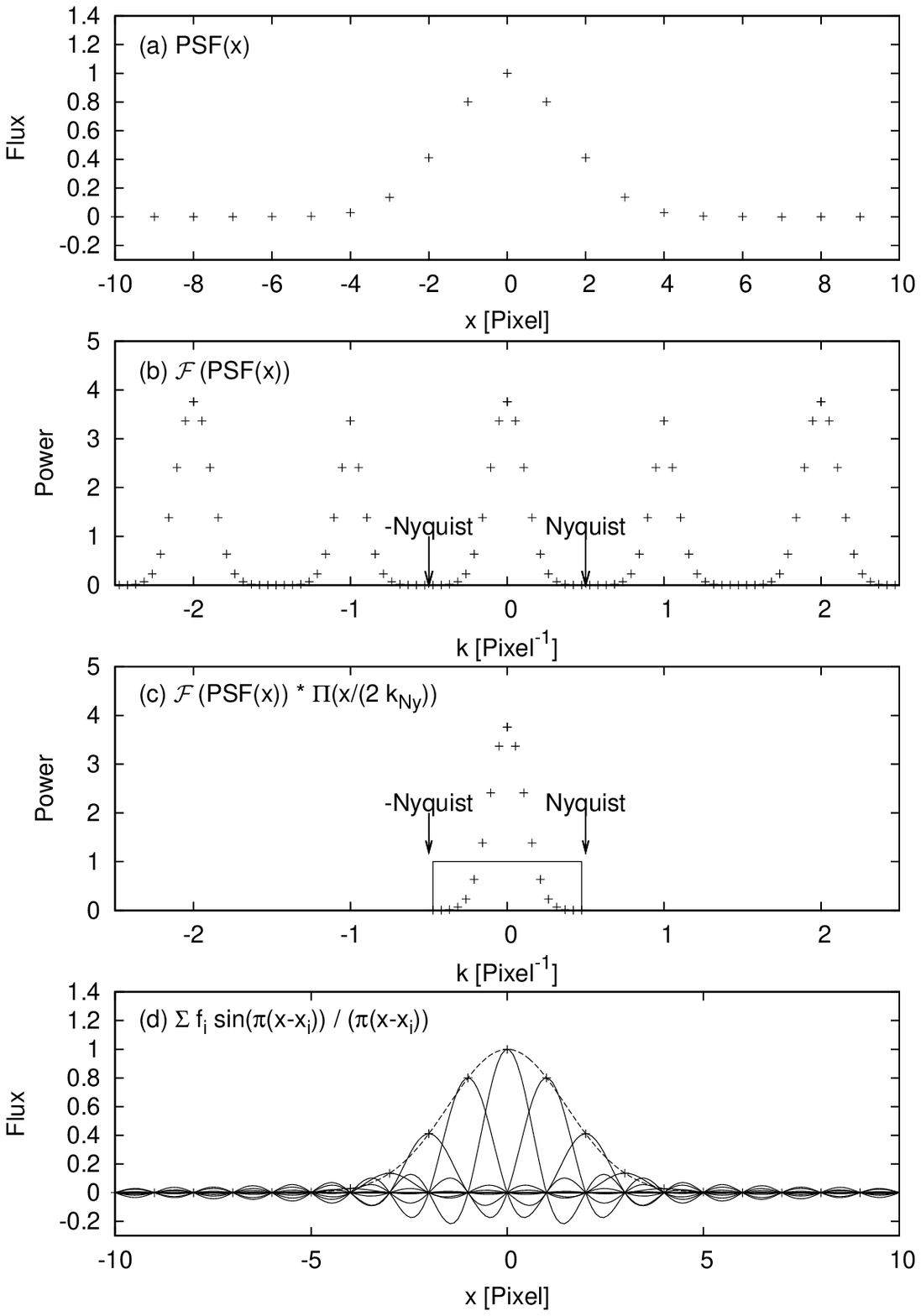}
    \caption
    {
      The sinc interpolation of a one-dimensional PSF.  A
      discretely sampled PSF (a) is transformed with a discrete
      Fourier transform (b), and truncated at the Nyquist wave number
      (c).  This truncation is the product of a boxcar and the
      wave-number coefficients, and is therefore represented by the
      convolution of a $\sinc$ (the Fourier transform of a boxcar)
      with the sampled points in real space (d).  If the sampled
      function is truly band limited, the $\sinc$ representation shown
      as a dashed line in (d) is exact.
    }
    \label{fig:samplingtheorem}
  \end{figure}
}

\newcommand{\figFluxerr} {
  \begin{figure}
    \centering
    \includegraphics[width=3in]{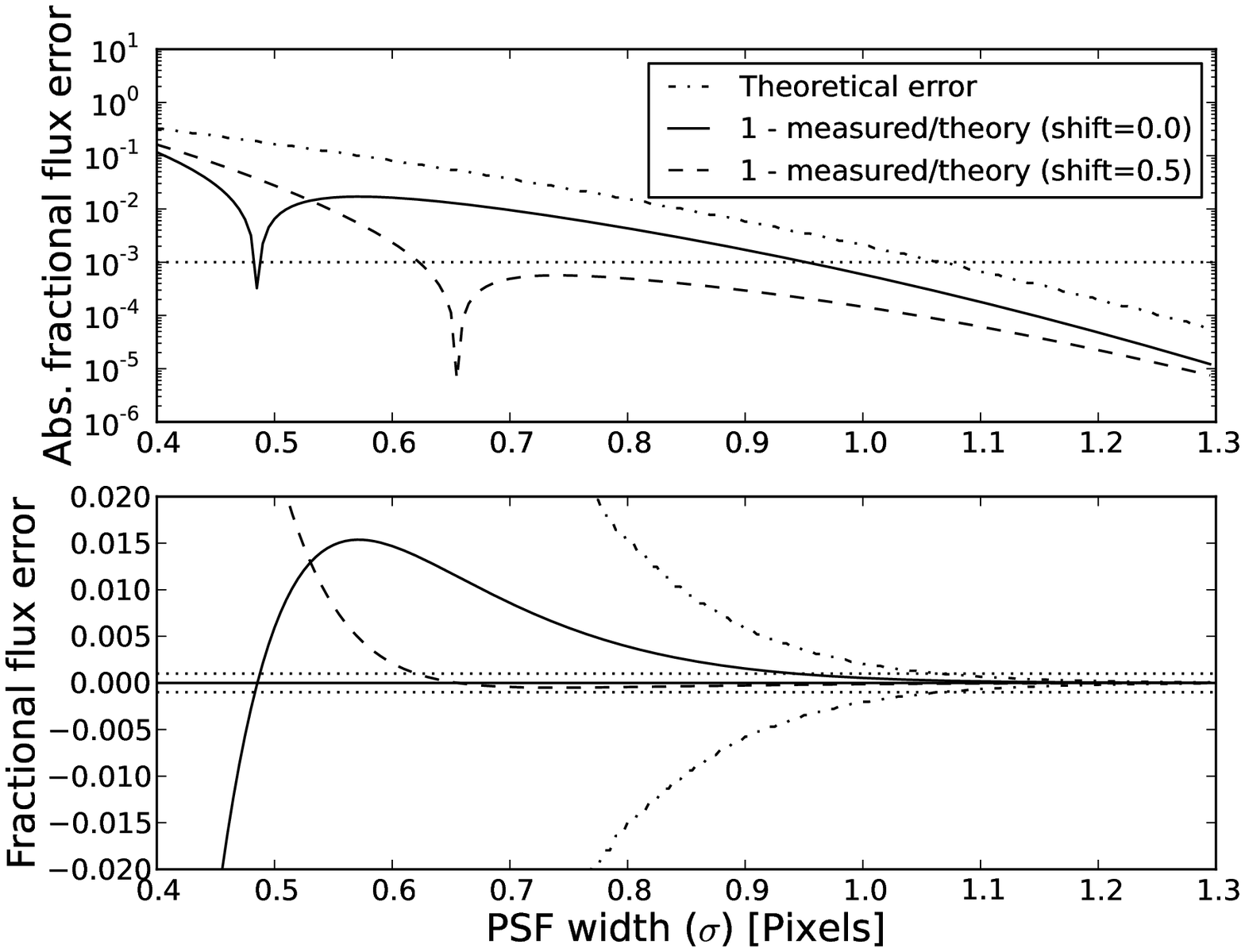}
    \caption
    { The flux error due to truncation/aliasing of wave numbers with
      $q > \pi$ as a function of PSF size in pixels for a double
      Gaussian (lower panel), with the absolute value shown in the
      upper panel.  The flux error is $1 - $measured/true, where
      `measured' refers to our $\sinc$-integration algorithm (aperture
      radius of 6 pixels), and `true' refers to the exact integral of
      the double Gaussian (sum of circular bivariate Gaussians having
      different amplitudes{\highlight (wing being 10\% of core)} and
      widths ($\sigma_{\mathrm{wing}}=2\sigma_{\mathrm{core}}$)).
      Curves are shown for a double Gaussian centred on an integer
      pixel value (solid), and one shifted by 0.5 pixels (dashed).
      Limits based on our conservative theoretical estimate are shown
      with dot-dashed lines.  The theoretical estimate is the limit
      for truncated flux, while the $\sinc$-integrated curves include
      aliased flux.  Dotted lines indicate a fractional error of
      0.001, or $\sim$1 mmag.  For the Gaussian PSFs, the flux error
      is $\ll 1$mmag provided sampling is $\sigma \gtrsim 1$
      pixel.{\highlight In cases of severe undersampling, the
        formalism of including the pixel response in the
        double-Gaussian PSF breaks down; therefore, values for small
        PSF widths ($\sigma_{\mathrm{core}} < 0.4$ pixels) are not
        shown.}  }
    \label{fig:fluxerr}
  \end{figure}
}

\newcommand{\figFootprint} {
  \begin{figure}
    \centering
    \includegraphics[width=3in]{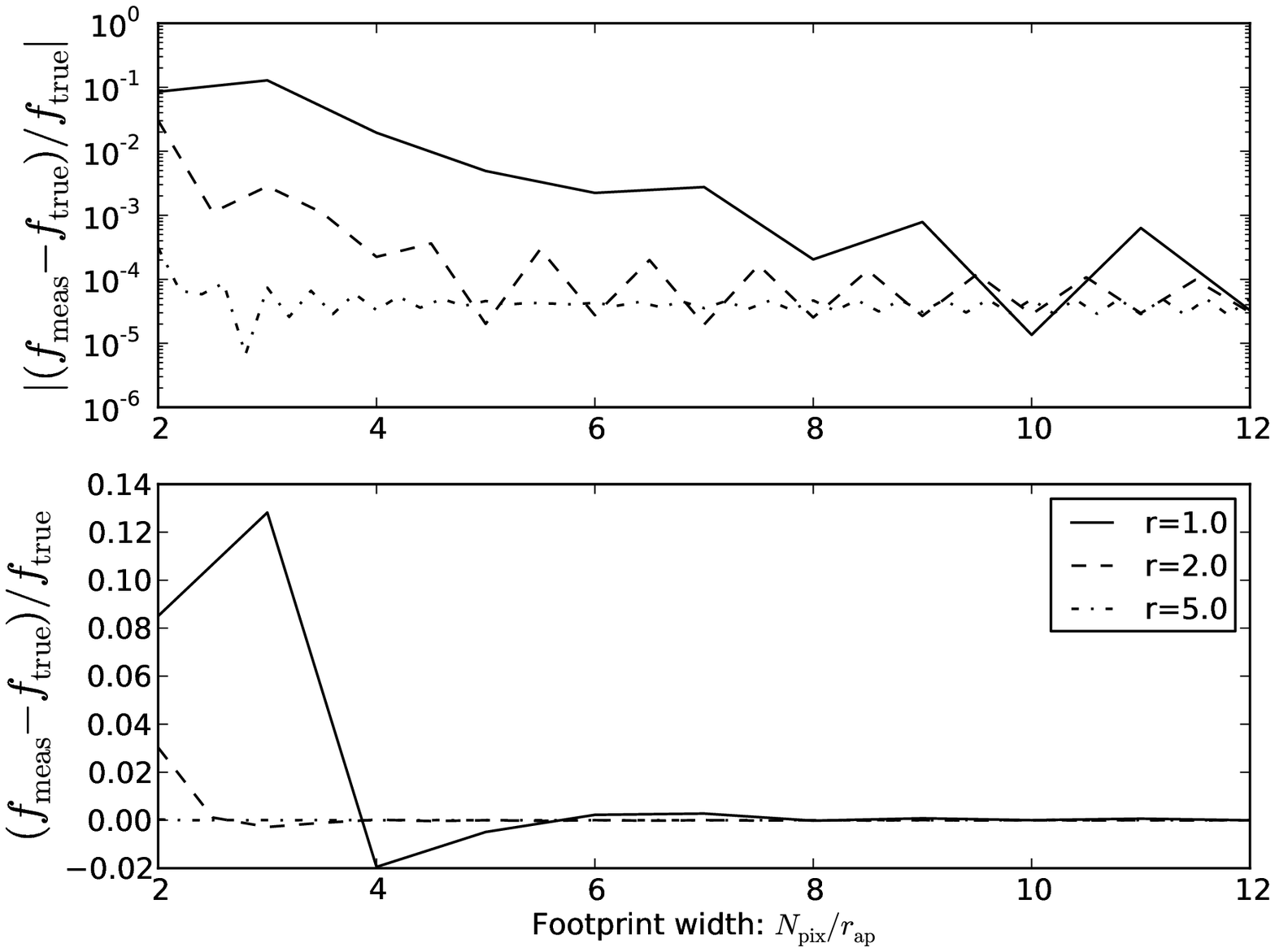}
    \caption
    {The flux error occuring at different footprint sizes (lower
      panel).  Absolute values are shown on a semilog scale (upper
      panel) to allow small values to be distinguished.  The flux
      error for a well-sampled ($\sigma=1.2$ pixel) double-Gaussian
      PSF is influenced by the width of the footprint used to compute
      the aperture flux.  The pixels outside the aperture carry
      weight, and the footprint must be larger than the aperture to
      include them.  Results for three aperture radii are shown:
      $r$=1.0, 2.0, and 5.0.  Footprint widths are normalized by the
      aperture radius.  Footprint widths smaller than two aperture
      radii would explicitly exclude a portion of flux and were not
      considered.  The error is large only when small apertures are
      used with small footprints.  A footprint having a 4 pixel border
      around the aperture is sufficient to achieve $<0.001$ fractional
      flux error for even the smallest aperture.  }
    \label{fig:footprint}
  \end{figure}
}

\newcommand{\figWijcirc} {
  \begin{figure}
    \centering
    \includegraphics[width=3in]{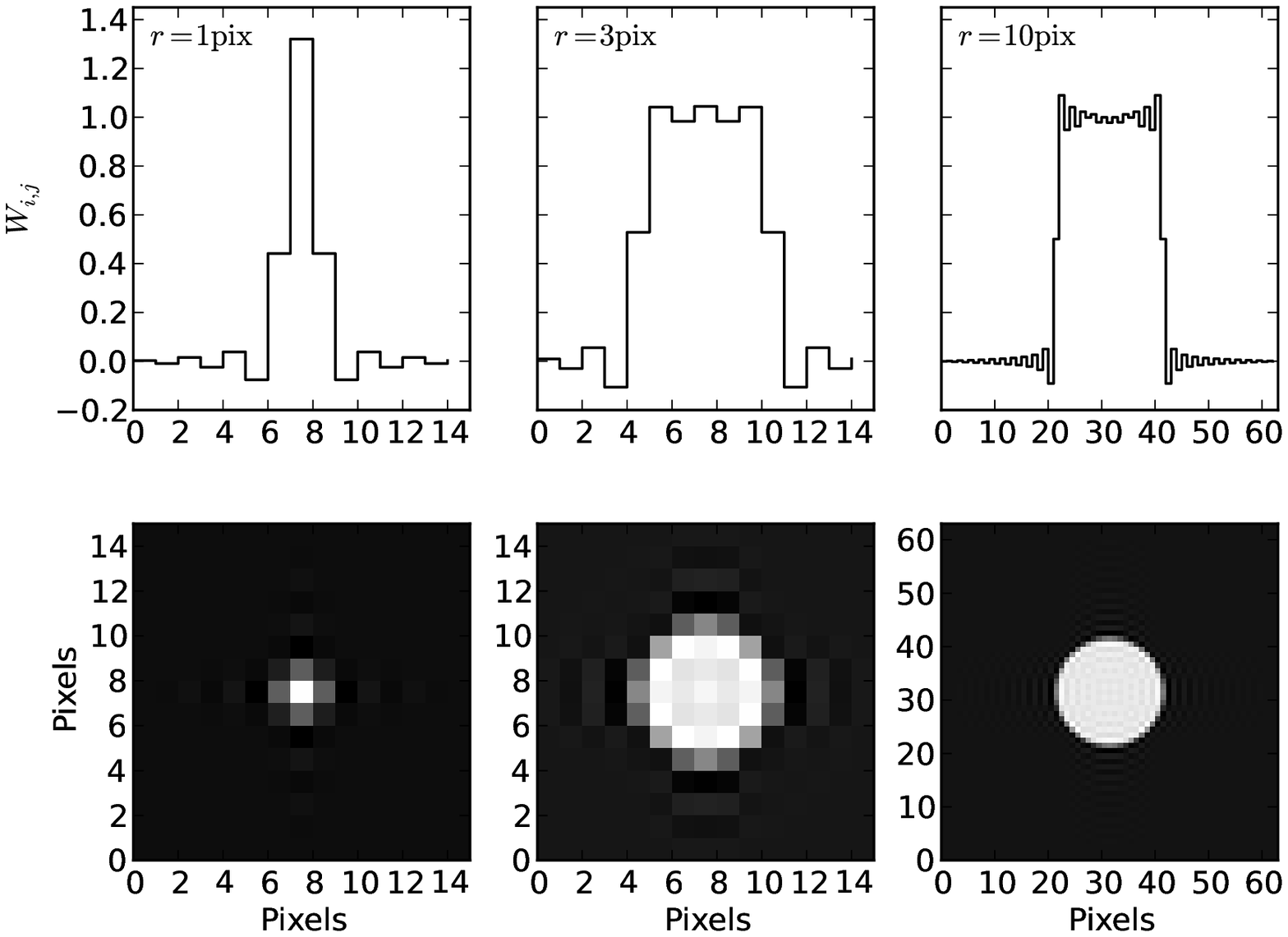}
    \caption
        { The $w_{ij}$ coefficients shown as images (bottom), and
          cross-sections through the middle row (top) for apertures with
          radii $r=1$ pix (left), $r=3$ pix (middle), and $r=10$ pix
          (right).  }
    \label{fig:wijcirc}
  \end{figure}
}

\newcommand{\figNea} {
  \begin{figure}
    \centering
    \includegraphics[width=3in]{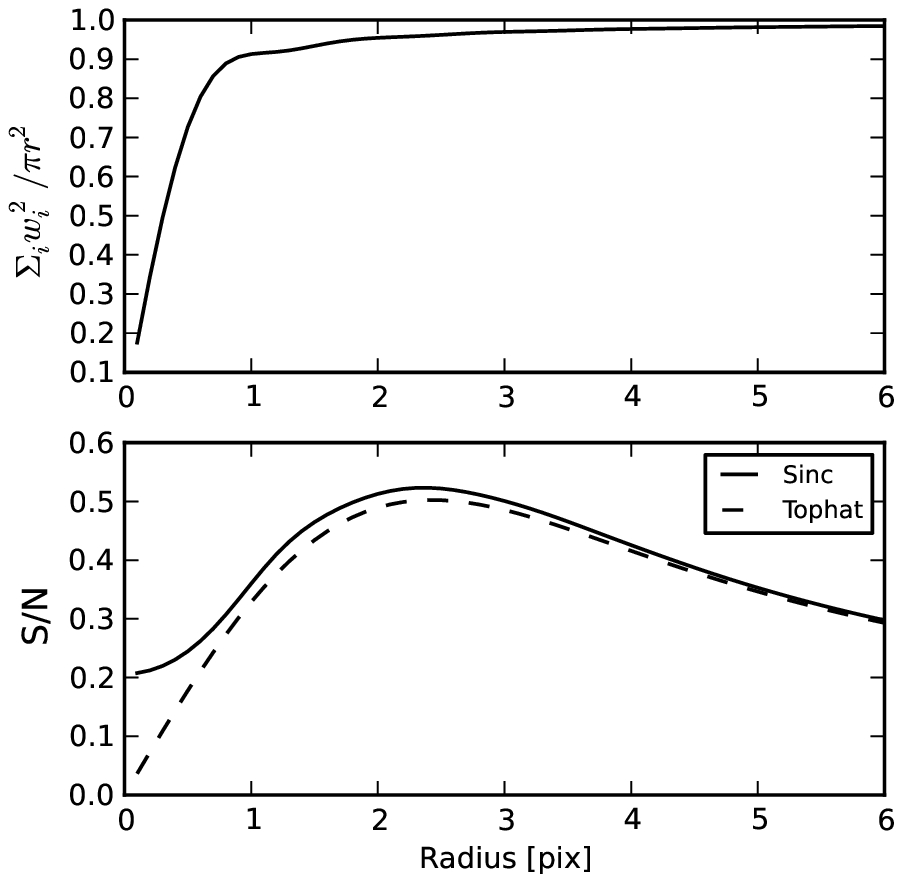}
    \caption{ The noise effective area ($\sum_i w_i^2 / \pi r^2$, top
      panel), and the signal to noise ratio (bottom panel) for
      circular sinc apertures as functions of aperture radius in
      pixels.  For large apertures, the noise effective area
      approaches 1 and S/N approaches that of a unity-weighted
      `tophat' aperture (shown dashed). For small radii, the noise
      effective area is smaller for the sinc aperture, and the S/N
      approaches a constant rather than zero.  The S/N curve is based
      on a double-Gaussian PSF with $\sigma_{\mathrm{core}}$=1.5
      pixel.  Optimal S/N for both methods is near
      $\sim1.6\sigma_{\mathrm{core}}$.}
    \label{fig:nea}
    \end{figure}
}

\newcommand{\figWijannul} {
  \begin{figure}
    \centering
    \includegraphics[width=3in]{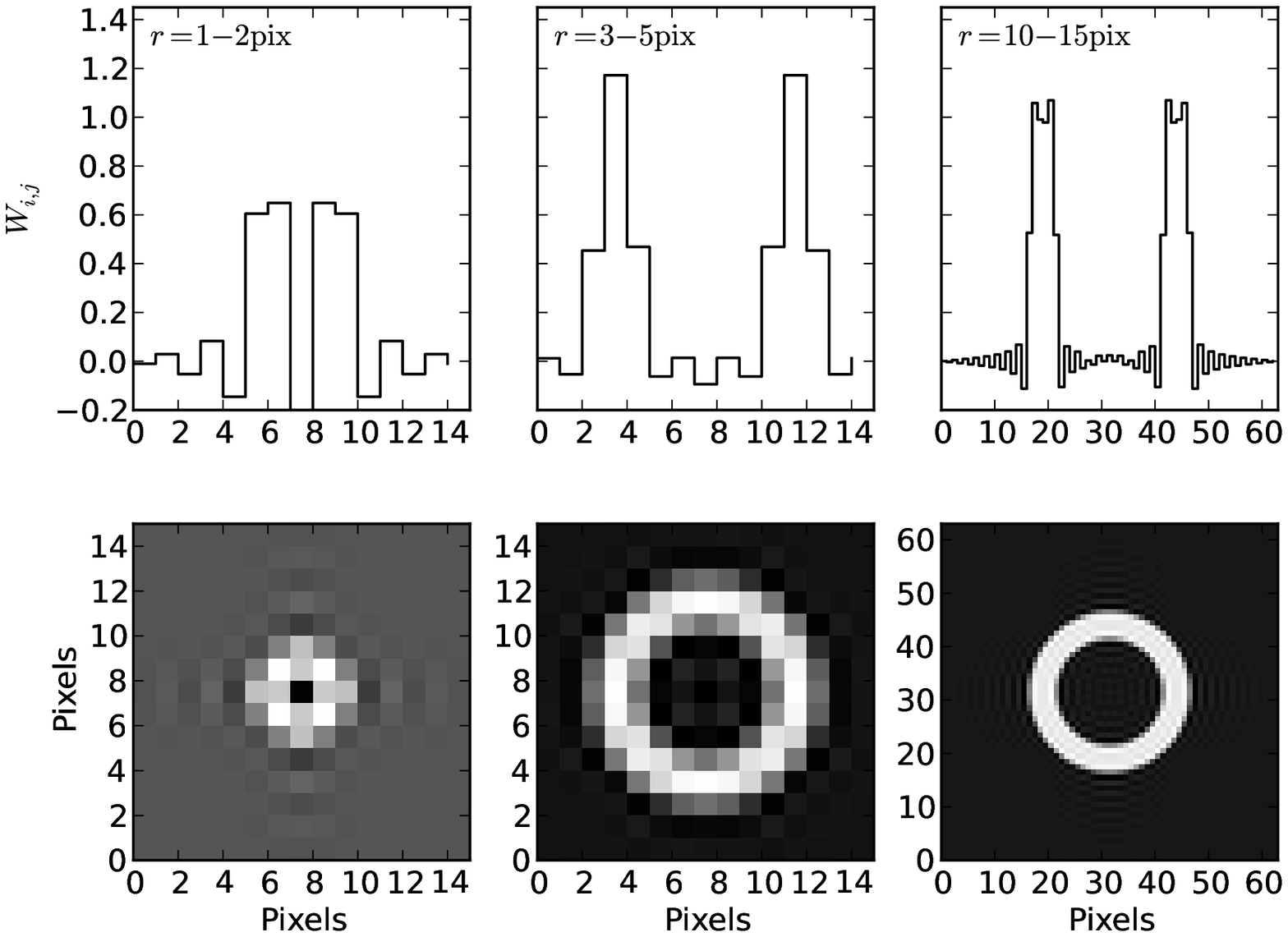}
    \caption
        { The $w_{ij}$ coefficients shown as images (bottom), and
          cross-sections through the middle row (top) for annular
          apertures with inner/outer radii $r=1-2$ pix (left),
          $r=3-5$ pix (middle), and $r=10-15$ pix (right).  }
    \label{fig:wijannul}
  \end{figure}
}

\newcommand{\figWijellip} {
  \begin{figure}
    \centering
    \includegraphics[width=3in]{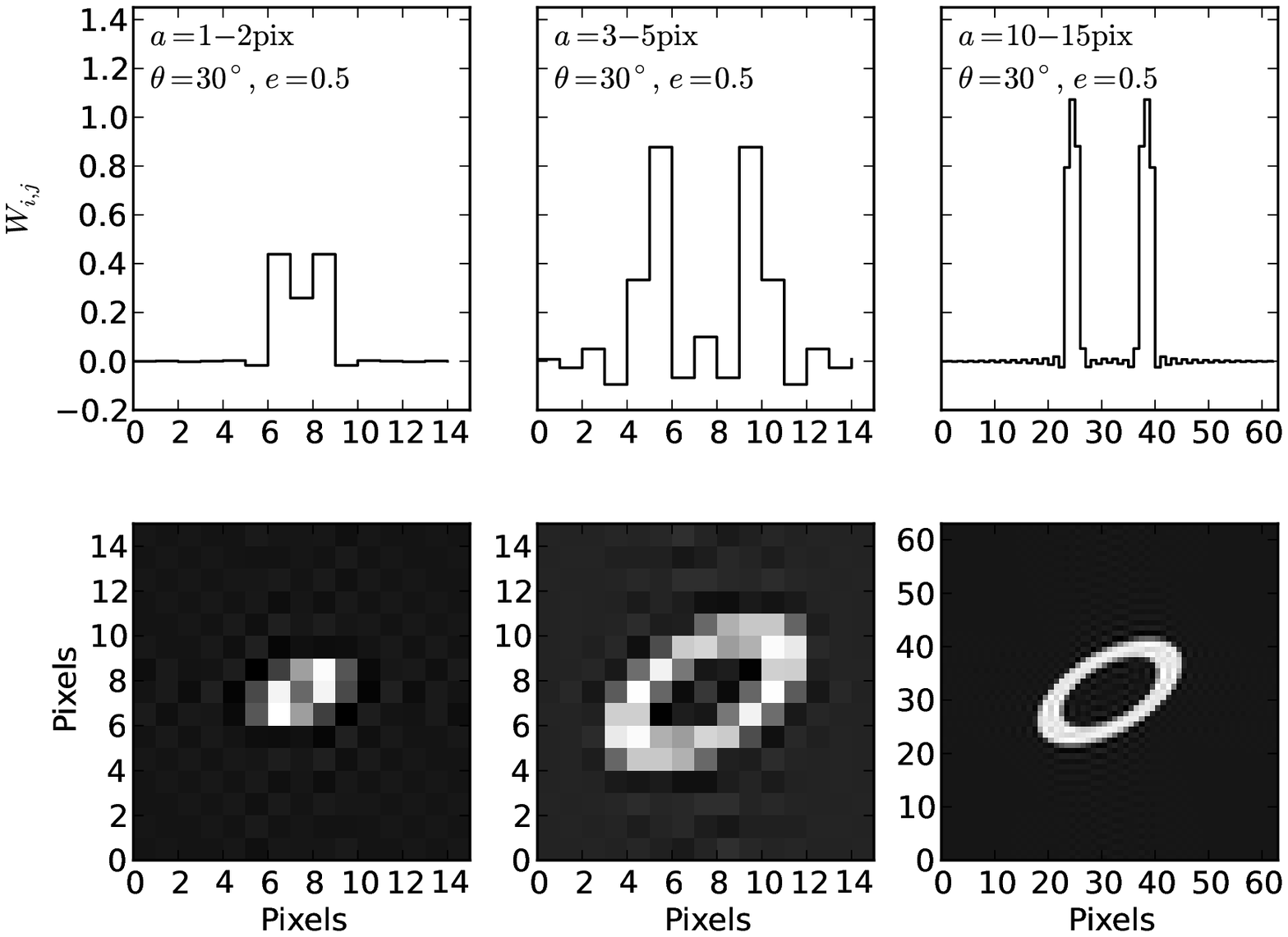}
    \caption
        { The $w_{ij}$ coefficients shown as images (bottom), and
          cross-sections through the middle row (top) for annular
          elliptical apertures with inner/outer semi-major axes
          $a=1-2$ pix (left), $a=3-5$ pix (middle), and $a=10-15$ pix
          (right).  Ellipses are rotated by $\theta=30^\circ$ and
          have ellipticity $e = 1 - b/a = 0.5$, where $b$ and $a$ are
          the semi-minor and semi-major axes, respectively. }
    \label{fig:wijellip}
  \end{figure}
}

\newcommand{\figGrowthcurve}{
  \begin{figure}
    \centering
    \includegraphics[width=3in]{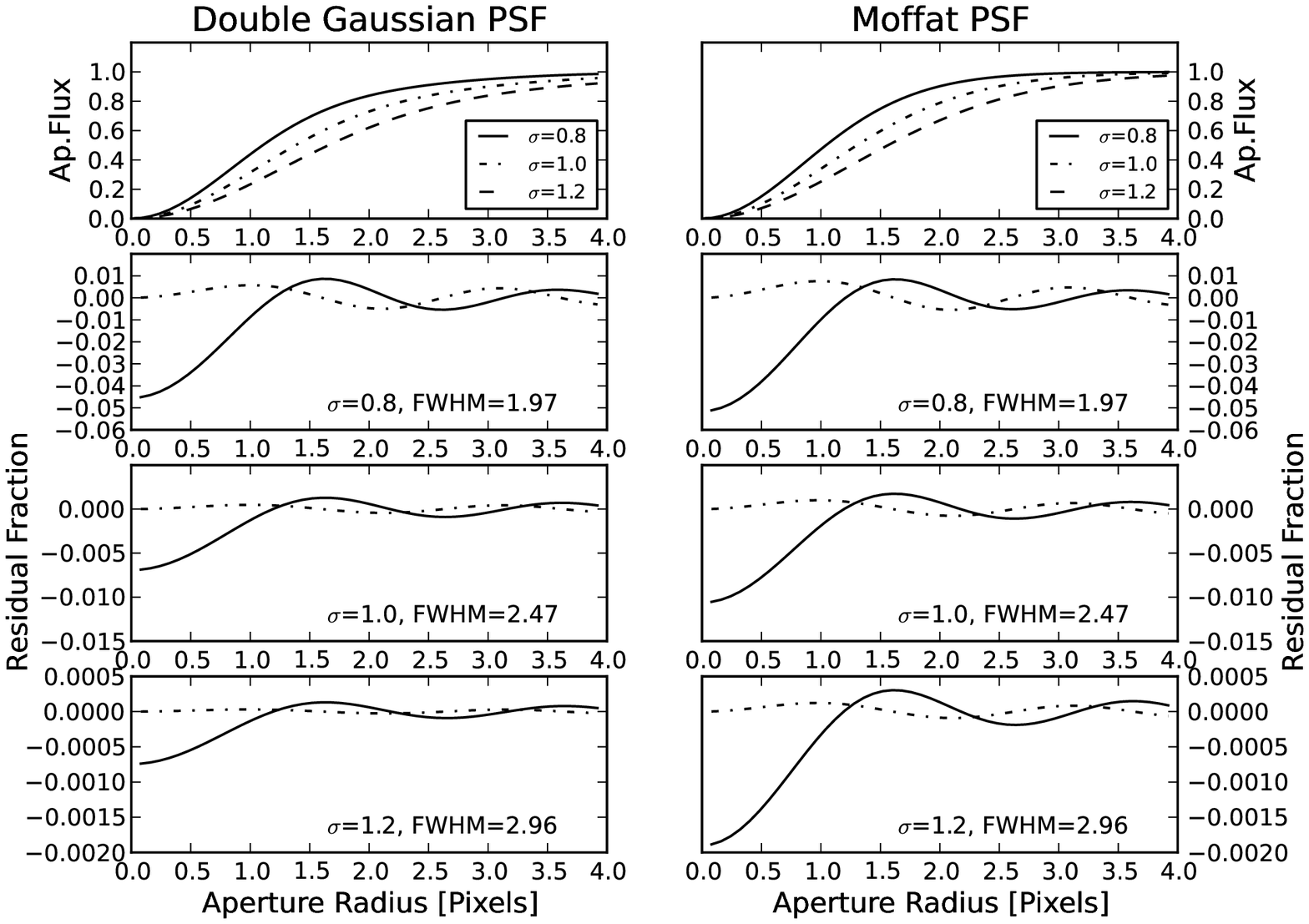}
    \caption
    {\highlight Measured growth curves and residuals with respect to
      computed theoretical fluxes.  Left and right panels show values
      for double-Gaussian and Moffat PSFs, respectively.  PSF widths
      of $\sigma$=0.8, 1.0, and 1.2 pixels were used to evaluate
      under-, critical-, and over-sampling.  The differences between
      the measured and theoretical curves are too small to be seen
      directly, but are shown as residuals in the three lower panels.
      Residuals shown with a dashed line are the minimum observed, and
      correspond to no sub-pixel shift.  Those shown with a solid line
      are the maxima, and correspond to a half-pixel shift in both
      $x$ and $y$ coordinates.  The curves are very similar in
      structure, but differ dramatically in amplitude.  The method is
      clearly unsuitable for sub-sampled ($\sigma < 1.0$ pixel) PSFs,
      but performs extremely well for $\sigma \gtrsim 1.2$ pixel.}
    \label{fig:growthcurves}
  \end{figure}
}

\newcommand{\figSexWellSampled} {
  \begin{figure}
    \centering
    \includegraphics[width=3in]{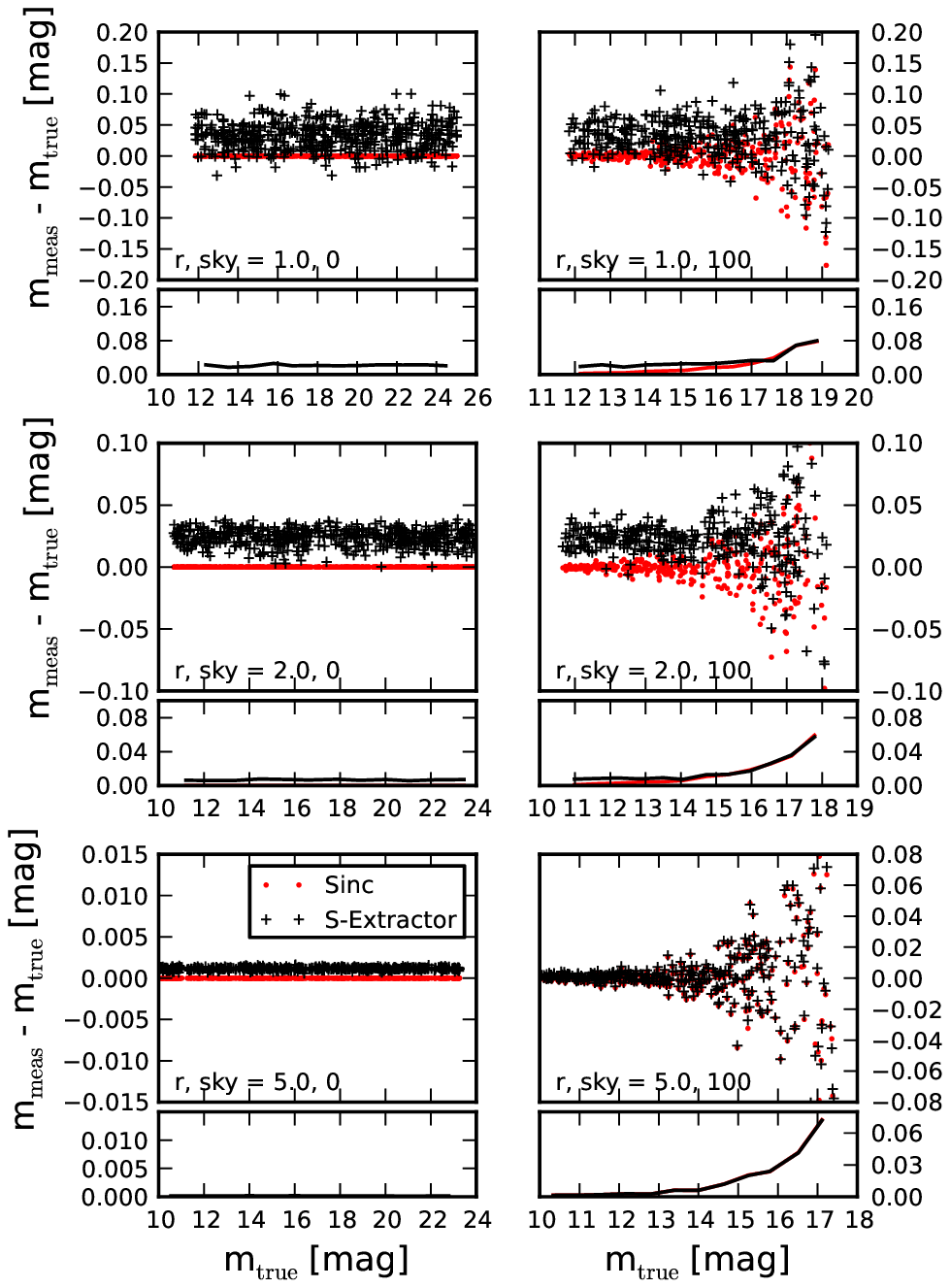}
    \caption
    { Magnitude error versus known magnitude for sinc (dots) and
      SExtractor ($+$ symbols) aperture magnitudes of {\itshape
        well-}sampled PSFs.  A lower sub-plot shows the standard
      deviation of points in bins with width 0.5 mag.  Well-sampled
      double-Gaussian PSFs ($\sigma_{\mathrm{core}}$=1.5 pixel) were
      added to $24\times24$ pixel images and were measured with both
      algorithms.  Aperture radii of $r$=1.0 (top), 2.0 (middle), and
      5.0 (lower) pixels were used.  Left panels show results for
      which no noise was added to the double-Gaussian PSFs, and right
      panels show results which included Poisson noise and the
      addition of a 100 count sky level.  For apertures with
      $r\gtrsim$5.0 pixels, performance is similar for both
      algorithms.  For smaller apertures, the sinc method outperforms
      in both precision and accuracy. }
    \label{fig:sexwellsampled}
  \end{figure}
}

\newcommand{\figSexUnderSampled} {
  \begin{figure}
    \centering
    \includegraphics[width=3in]{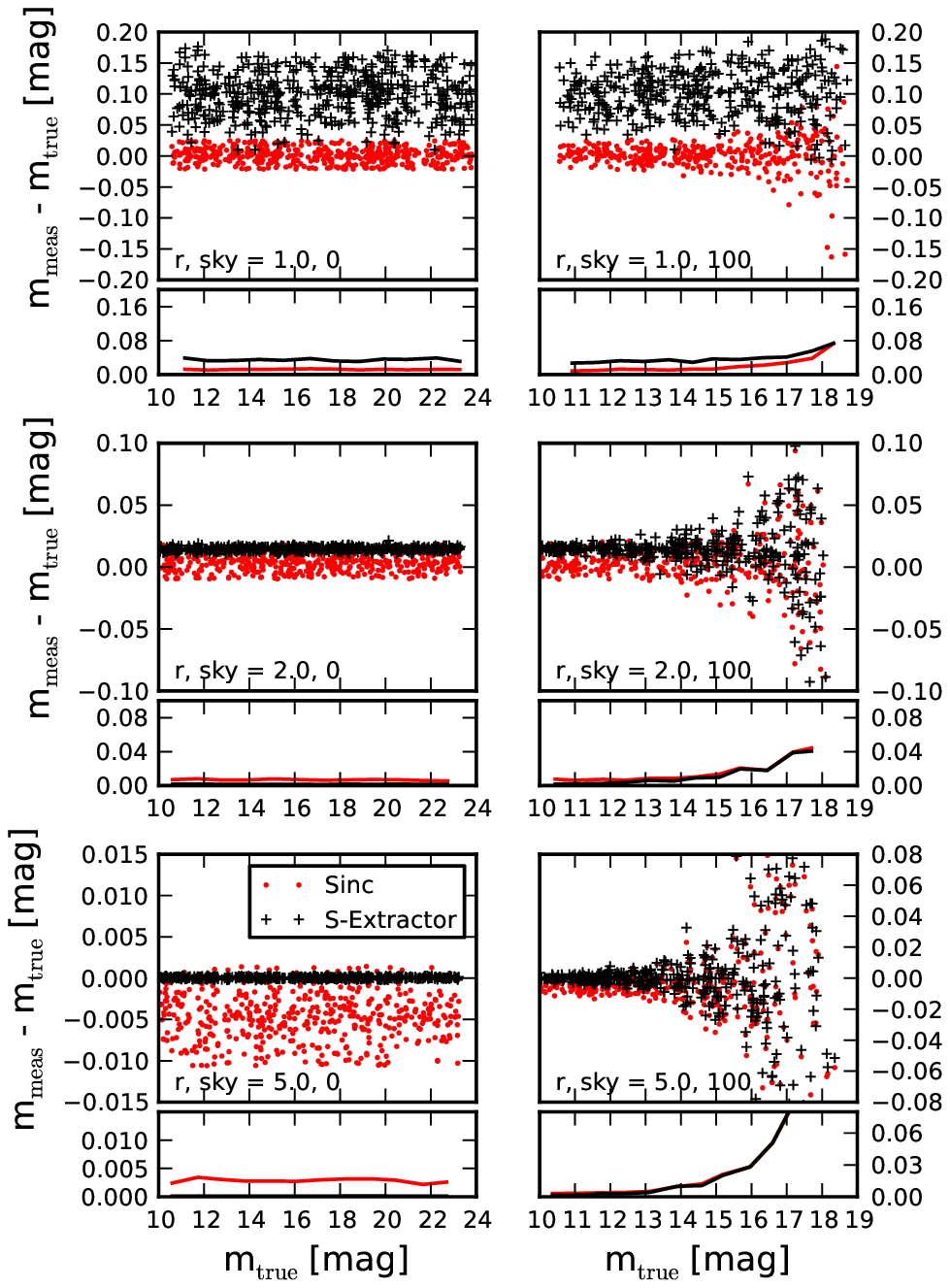}
    \caption
    { Magnitude error versus known magnitude for sinc (dots) and
      SExtractor ($+$ symbols) aperture magnitudes of {\itshape
        under-}sampled PSFs.  The panels are the same as those
      described in Figure~\ref{fig:sexwellsampled}.  In this case,
      {\itshape under-}sampled (0.7 pixel) double-Gaussian PSFs were
      used.  For large apertures the SExtractor magnitudes are more
      precise and more accurate than the sinc magnitudes. For small
      apertures, the sinc magnitudes are more precise and more
      accurate, becoming the better option for apertures with
      $r\lesssim$1.0 pixel.}
    \label{fig:sexundersampled}
  \end{figure}
}

\newcommand{\figLsst} {
  \begin{figure}
    \centering
    \includegraphics[width=3in]{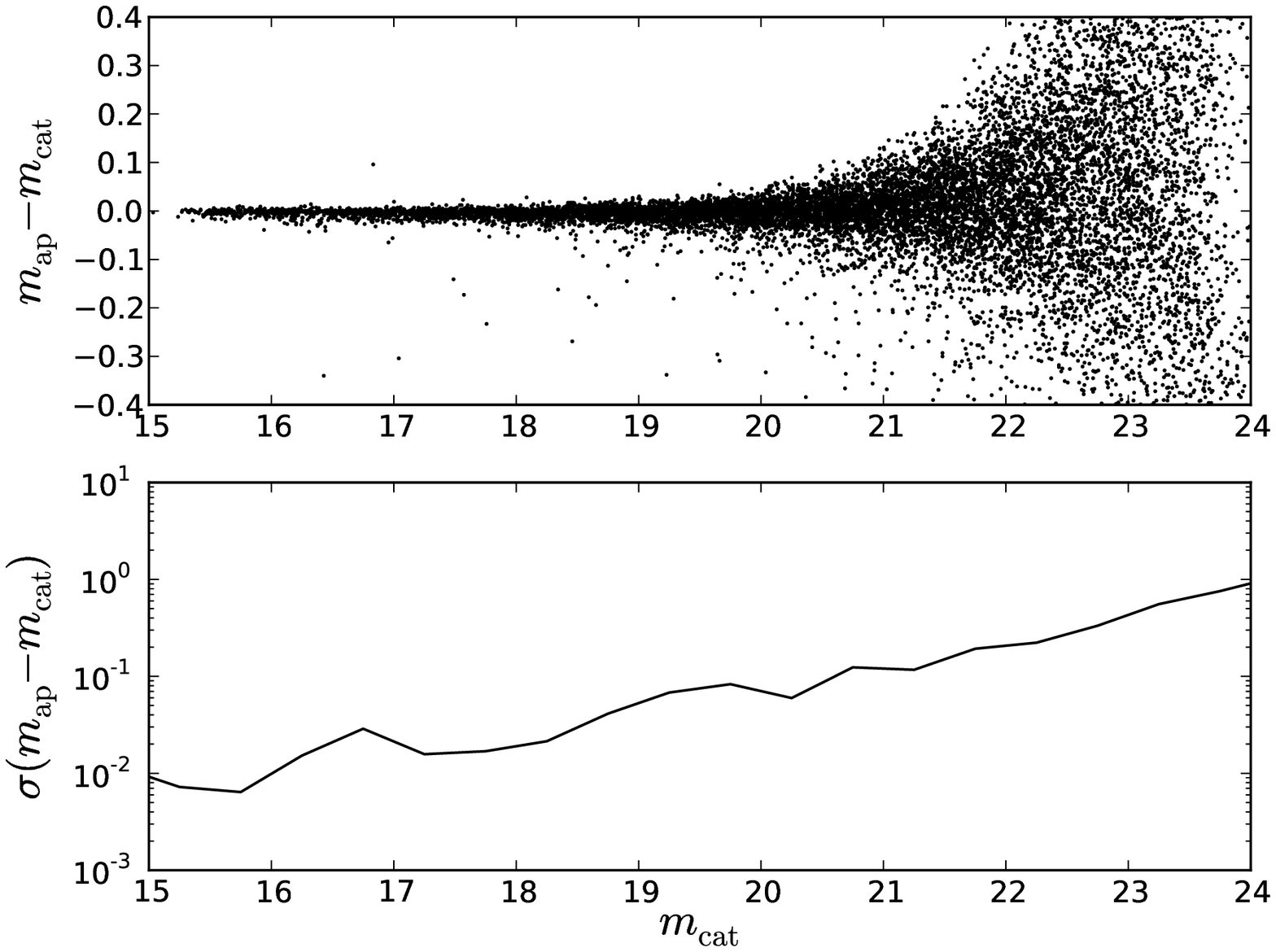}
    \caption
        {{\highlight Sinc aperture photometry performed on a simulated
            LSST $i'$ image \citep[][in preparation]{peterson13}.
            For clarity, only 1/4 of the values from a full frame are
            shown.  The upper panel shows
            $m_\mathrm{ap}-m_\mathrm{cat}$ versus $m_\mathrm{cat}$,
            and the lower panel shows the standard deviation of the
            upper panel data in 0.5 mag bins.  The aperture radius was
            7 pixels (1.4\arcsec), and magnitudes are compared to the
            exact known values (i.e. $m_\mathrm{cat}$) used as input
            for the simulator.}  }
        \label{fig:lsst}
  \end{figure}
}

\ifthenelse{\boolean{astroph} \or \boolean{mn2e}}{
  \newcommand{\middfig}[1]{{#1}}
  \newcommand{\tailfig}[1]{{}}
}{
  \newcommand{\tailfig}[1]{{#1}}
  \newcommand{\middfig}[1]{{}} 
}

\ifthenelse{\boolean{astroph}} {
  
}{}

\newcommand{\myemail}{bick@astro.princeton.edu}

\begin{document}

\ifthenelse{\boolean{mn2e}}{
  \title[Precise Aperture Photometry]{An Algorithm for Precise Aperture Photometry of
    Critically Sampled Images}
  \author[S.J. Bickerton]{S.J. Bickerton\thanks{E-mail: \myemail}, and R.H. Lupton\\
    Department of Astrophysical Sciences, Princeton University, Princeton, NJ
    08544}
  \date{\today}
  \pagerange{\pageref{firstpage}--\pageref{lastpage}} \pubyear{2013}
  \maketitle
}{
  \shorttitle{Precise Aperture Photometry}
  \shortauthors{Bickerton and Lupton}

  \title{An Algorithm for Precise Aperture Photometry}
  \author{S.J. Bickerton, and R.H. Lupton}
  \affil{Department of Astrophysical Sciences, Princeton University,
    Princeton, NJ 08544}
  \email{bick@astro.princeton.edu}
  \email{rhl@astro.princeton.edu}
}

\label{firstpage}

\begin{abstract}
  We present an algorithm for performing precise aperture photometry
  on critically sampled astrophysical images.  The method is intended
  to overcome the small-aperture limitations imposed by
  point-sampling.  Aperture fluxes are numerically integrated over the
  desired aperture, with sinc-interpolation used to reconstruct values
  between pixel centers.  Direct integration over the aperture is
  computationally intensive, but the integrals in question are shown
  to be convolution integrals and can be computed $\gtrsim
  10000\times$ faster as products in the wave-number domain.  The
  method works equally well for annular and elliptical apertures and
  could be adapted for any geometry. A sample of code is provided to
  demonstrate the method.
\end{abstract}

\ifthenelse{\boolean{astroph} \or \boolean{aastex}} {
}{}

\section{Introduction}
\label{intro}

In aperture photometry, one seeks to measure the flux which would have
been received had the light passed through a physical aperture (e.g. a
photomultiplier tube).  An {\itshape aperture flux} in an
astrophysical image is typically measured as the sum of the pixel
fluxes for pixels near a given source.  The region summed over is the
`aperture', and may contain whole and/or partial or weighted pixels.
Circular apertures are common for astrophysical measurements, and are
implemented in some form in most standard software packages, including
IRAF \citep{tody86}, DAOPHOT \citep{stetson87}, and SExtractor
\citep{bertin96}.

Aperture fluxes are most commonly used in calibration when
constructing a `growth curve'. A growth curve plots the enclosed flux
within an aperture as a function of its radius, and is essential for
calibrating fluxes measured through an optimal (typically small)
aperture to bright isolated standard stars which must be measured with
a large aperture to include all flux.  Also, a number of commonly used
flux measurements are aperture fluxes computed for scientifically
interesting aperture shapes or sizes.  Examples include Kron fluxes
\citep{kron80} and Petrosian fluxes{\highlight \citep{petrosian76}}.

For aperture radii comparable to the pixel dimensions, if no partial
pixels are used, the `circular' aperture can be quite non-circular
(e.g. a $1\times1$ or $2\times2$ square, a cross, etc.).  As the flux
may change significantly across a pixel, the use of geometric partial
pixels (e.g. including 0.5$\times$flux for a pixel with half of its
area enclosed in the aperture) will not accurately represent the flux
contained in the corresponding fraction of the pixel.  To maintain a
truly constant aperture geometry (circular or otherwise), the flux
must be interpolated between pixels and integrated.

Here we describe a straightforward aperture photometry algorithm,
which under normal observing conditions preserves circular aperture
geometry perfectly and remains well-defined to arbitrarily small
radii. For the method to be accurate, the pixels must critically
sample the point-spread function (PSF); i.e. the spatial frequency of
the pixels must be at least twice the highest spatial frequency
present in the PSF. The method works equally well for annular and
elliptical apertures, and could be adapted for any desired shape.  The
ability to handle elliptical apertures, in particular, makes it useful
for computing Kron and elliptical Petrosian fluxes.  It is implemented
in the image processing pipeline under development for the Large
Synoptic Survey Telescope (LSST) \citep{ivezic05}.

To compute the flux that would have been received through a truly
circular aperture centred on a point source, we perform a continuous
reconstruction of the discretely sampled pixel fluxes
(Section~\ref{sec:sincmodel}) and integrate to yield the circular
aperture flux (Section~\ref{sec:apphot}).  The two-dimensional
integral is computationally intensive, but we use Fourier methods to
compute it quickly without loss of precision (Section~\ref{sec:fft}).
In Section~\ref{sec:otherapertures}, we modify the method to compute
annular and elliptical apertures; and in Section~\ref{sec:performance}
we evaluate the performance of the algorithm by comparing growth
curves for{\highlight model} point-spread functions (PSFs) to their
known values.  Results are discussed and summarized in
Sections~\ref{sec:discussion} and~\ref{sec:summary},
respectively.{\highlight Some additional calculations are included in
  Appendix~\ref{app:aliasedpower}, and} an example program (written in
Python) is included in Appendix~\ref{app:code}.

\section{Reconstructing a Discretely Sampled PSF}
\label{sec:sincmodel}

A telescope produces a continuous function and delivers it to a
detector, which convolves with the pixel response and point-samples it.
In our technique, we use the sampling theorem to recover the
continuous function, and measure it.  Our only assumption is that the
function must be band-limited.  In this context, this implies that the
Fourier transform of the PSF has negligible power above the
Nyquist wave number which corresponds to the pixel sampling.

The sampling theorem states that a discretely sampled continuous
function can be completely represented by its samples.  In
one dimension

\begin{equation}
  \label{eq:sampletheorem}
  p(x) = \sum_i\ p_i\ \frac{\sin(\pi (x - x_i))}{\pi (x - x_i)}.
\end{equation}

Equation~\ref{eq:sampletheorem} can be used to interpolate $p(x)$
given the discrete samples $p_i$.  As the $\sin(x)/x$ term is a
$\sinc$ function, this is known as $\sinc$ interpolation.
Figure~\ref{fig:samplingtheorem} demonstrates how the sinc
interpolation of a one-dimensional PSF can be understood in terms of
Fourier methods.  If the sampled function is truly band-limited (no
power above the Nyquist wave number), the $\sinc$ interpolation is an
exact representation of the underlying continuous function.  The proof
of the sampling theorem can be found in the original work by
\citet{shannon49}.

\middfig{\figSamplingtheorem}

The sinc decomposition of a two-dimensional function has the form

\begin{equation}
  \label{eq:sampletheorem2d}
  p(x,y) = \sum_j \sum_i\ p_{ij}\ \frac{\sin(\pi (x - x_i))}{\pi (x - x_i)}\ \frac{\sin(\pi (y - y_j))}{\pi (y - y_j)}.
\end{equation}

\subsection{Testing the Band-Limit of a Gaussian PSF}
\label{ssec:bandlimittest}

The requirements for the $\sinc$ reconstruction to be valid are that
the PSF be band limited, and that an infinite sequence of values are
available.  However, in both cases more practical limits are
sufficient.  Here we test the first assumption (the band limit) in the
context of reconstructing a continuous PSF.  Requirements on the
extent of the data are dealt with in Section~\ref{ssec:footprint}.

{\highlight The validity of the band-limit assumption was verified by evaluating
the error in integrated flux as a function of the sampling interval.
This was estimated analytically with a circular bivariate
single-Gaussian PSF model, and also measured numerically with a
double-Gaussian PSF.  A PSF is typically well-modeled by a double
Gaussian (a sum of two bivariate Gaussians, one modeling the PSF core
and the other modeling the wings). The single-Gaussian model used for
the analytic estimate is intended to test the core Gaussian of a
double-Gaussian PSF.  If the `band limit' criterion is satisfied for
the narrower Gaussian, it would easily be satisfied for the larger one
representing the wings.

For our analytic model, the calculation of the fractional error due to
aliased power in a sampled Gaussian is a straightforward exercise, and
is presented in Appendix~\ref{app:aliasedpower}.  The resulting
upper-limit estimate for the fractional flux error is

\begin{equation}
\frac{\mathrm{error}}{\mathrm{total\ flux}} = \frac{e^{-\pi^2\sigma^2/2}}{2\pi^{1/2}\sigma}.
\label{eq:fluxerr}
\end{equation}

} 

To verify the method more rigorously, it was also tested numerically
with known PSFs.  For the numerical tests, the so-called `double
Gaussian' is a{\highlight reasonable} approximation to a typical
observed PSF.  Our double-Gaussian test PSF has core and wing
components,{\highlight with the wing component having 10\% of the
  core's amplitude and a width of $\sigma_{\mathrm{wing}} =
  2\sigma_{\mathrm{core}}$. A double Gaussian also has the advantage
  that it can be integrated analytically to provide an exact known
  flux for comparison to the $\sinc$-integrated value, and it was
  therefore used for these numerical tests.

  The response of the pixel is considered to be a part of the PSF
  structure (i.e. the continuous PSF produced by the telescope is
  convolved with a pixel and then point-sampled by the detector), and
  the double Gaussian model used in the numerical tests was directly
  point-sampled. As convolving a Gaussian with a pixel is {\itshape
    very} close to convolving with a another Gaussian having
  $\sigma=1/\sqrt{12}$\footnote{The RMS width of a unit boxcar
    function (i.e. a pixel) is $1/\sqrt{12}$.}, the resulting function
  remains very close to a Gaussian. }

Both theoretical (equation~\ref{eq:fluxerr}) and numerically measured
errors are shown as a function of the PSF size in pixels (i.e., the
sampling) in Figure~\ref{fig:fluxerr}.{\highlight An aperture with a
  radius of 6 pixels was used.}  For the numerical tests, sub-pixel
centroid shifts were applied to verify the method for PSF profiles
that are not centred at integer pixel positions, and representative
curves corresponding to shifts of $(\delta_x,\delta_y)$ = $(0.0,
0.0)$; and $(0.5, 0.5)$ are presented.{\highlight Provided the PSF's
  $\sigma_{\mathrm{core}}$ is larger than $\sim$1 pixel (corresponding
  to a FWHM $\gtrsim2.5$ pixels for a double Gaussian)}, fractional
flux errors due to undersampling are $\ll$ 0.001, or $\sim$1 mmag.

\middfig{\figFluxerr}

It is tempting to assume that any error introduced through aliasing
would be systematic and correspond to a fractional excess or deficit
of flux.  Such an error could be corrected by calibrating the measured
fluxes against standard stars.  However, Figure~\ref{fig:fluxerr}
shows that the errors are different for the $(\delta_x,\delta_y)$ = $(0.0,
0.0)$; and $(0.5, 0.5)$ pixel shifts.  In general, the error for each
measured source depends on the sub-pixel shift of the PSF and cannot
be corrected through calibration.  Further, galaxies and other
non-point-sources will be composed of different spatial frequencies
and their errors will be different from those of point sources and
from other galaxies.

If the method is to be used in the construction of a growth curve
(flux measurements at a series of progressively larger apertures), the
band-limited nature of the PSF must be taken seriously as the error is
a function of radius and can be large for undersampled
PSFs.{\highlight For additional information, see
  Section~\ref{ssec:growthcurve} where the performance of the method
  is tested directly by measuring growth curves.}

With these caveats in mind, a double-Gaussian PSF {\itshape is} band
limited for the pixel scales{\highlight and aperture sizes} commonly
used in astronomy ($\sigma > 1$ pixel, or FWHM $> 2.5$ pixels).

\section{Computing Aperture Flux}
\label{sec:apphot}

With the discrete pixel fluxes, $p_{ij}$, interpolated to form a
continuous function, $p(x, y)$ (see equation~\ref{eq:sampletheorem2d}),
a precise aperture flux, $f_A$, can be obtained by integrating over
the desired aperture $A$:

\begin{equation}
\label{eq:apflux}
  f_A = \iint\limits_A\ p(x,y)\ \dif x \dif y.
\end{equation}

The integration can be performed to arbitrary precision, but depending
on the structure of the aperture, and integrator used, it may be
computationally intensive (recall that the $p(x,y)$ function is a sum
of $\sinc$ functions -- one for each pixel).

Fortunately, the $p_{ij}$ terms in equation~\ref{eq:sampletheorem2d}
are constant and can be factored out of the integral in
equation~\ref{eq:apflux}, allowing the aperture flux to be expressed
as a weighted sum:

\begin{equation}
\label{eq:fluxsum_pq}
  f_A = \sum_j \sum_i\ w_{ij}\ p_{ij},
\end{equation}

\noindent where the weighting terms are

\begin{equation}
\label{eq:intsinc}
w_{ij} = \iint\limits_A \ \frac{\sin(\pi (x - x_i))}{\pi (x -
  x_i)}\ \frac{\sin(\pi (y - y_j))}{\pi (y - y_j)}\ \dif x\ \dif y.
\end{equation}

Thus, for each pixel we compute a corresponding weight, $w_{ij}$, in
the sum for the aperture flux.  The integral ($w_{ij}$) does not
depend on $p_{ij}$.  It can be computed once and shifted (see
Section~\ref{ssec:shift}) to be applied to different sources in a
given frame.  This is critical for efficiency.  Without the ability to
pre-compute the $w_{ij}$ coefficients, several numerical integrals
would need to be computed for each source being measured, and the
method would be prohibitively slow.

For a large circular aperture, pixels near the center have weight
$w_{ij}\approx 1$.  Near the edge of the aperture, values transition
from $\sim$1 to $\sim$0.  However, values outside the aperture do not
go completely to 0. Some non-zero weight remains even several pixels
outside the aperture radius.  Examples of $w_{ij}$ coefficients for
different-sized circular apertures are shown as images and
cross-sections in Figure~\ref{fig:wijcirc}.

\middfig{\figWijcirc}

\subsection{The Minimum Size Footprint}
\label{ssec:footprint}

As the $w_{ij}$ coefficients are non-zero outside the aperture radius,
they cannot be ignored and the footprint for the aperture (the
$n_x\times n_y$ image containing the $w_{ij}$ coefficients) must be
larger than the aperture itself.  To determine the required footprint
size, Gaussian PSFs were planted in images of varying size:
$2-12\times r_{\mathrm{ap}}$.  These were compared to the known values
for a double Gaussian analytically integrated over the same aperture.
A well-sampled double Gaussian ($\sigma_{\mathrm{core}}=1.2$ pixels)
was used with aperture radii of $r_{\mathrm{ap}}$=1.0, 2.0, and 5.0
pixels.  For small apertures ($<2$ pixel), errors can be as large as a
few percent if the footprint is too small.  For large apertures,
errors are negligible even for footprints only slightly larger than
the aperture radius.  Empirically, we find that with a 4 pixel border
around the aperture, errors are reduced to $<0.001$ for a well-sampled
PSF.  Figure~\ref{fig:footprint} shows the flux error as a function of
the width of the footprint.

\middfig{\figFootprint}

\subsection{The Effects of Noise}
\label{ssec:noise}

Poisson noise is always present in an astrophysical image and has a
flat power spectrum which is not band limited.  The sinc aperture flux
$f$ can be regarded as a simple weighted sum of pixel intensities,
$I_i$.

\begin{equation}
\label{eq:wsum}
f = \sum_i w_i I_i
\end{equation}

The variance of the sum, assuming a pixel variance $\sigma_i^2$ is

\begin{equation}
\label{eq:wsumvar}
\sigma_f^2 = \sum_i w_i^2 \sigma_i^2.
\end{equation}

Equations~\ref{eq:wsum} and~\ref{eq:wsumvar} remain true regardless of
the choice of weight values, and the variances of two candidate
weighting methods can be compared as a ratio.  For constant pixel
noise $\sigma_I$ (i.e. sky limited observations), the sinc aperture
compared to a top hat aperture having $w_i = 1$ gives a ratio of

\begin{equation}
  \label{eq:ratio}
  \sigma_{f_\mathrm{sinc}}^2 / \sigma_{f_\mathrm{tophat}}^2 = \sum_i w_i^2 / N.
\end{equation}

The number of pixels $N$ is the area of the comparison aperture, and
$\sum_i w_i^2$ can be regarded as a {\itshape noise equivalent area}
{\highlight \citep{king83}}, i.e. the area of a unity-weighted aperture which would
contribute the same variance to the measurement.  For the sinc
aperture, Figure~\ref{fig:nea} (upper panel) shows
equation~\ref{eq:ratio} versus aperture radius, with the denominator
replaced by the area of the circular aperture being measured.

\middfig{\figNea}

For large aperture radii, the noise equivalent area asymptotically
approaches 1, but it decreases for small apertures.  The noise in a
sinc flux is {\itshape less} than would be expected in a
unity-weighted `top hat' aperture!  The inclusion of pixels outside the
aperture radius might be expected to introduce additional noise, but
instead serves to reduce the overall variance.  The pixels are present
outside the aperture radius, but they carry less weight and the noise
is averaged over a larger sample of pixels.

Note that we have not considered the shape of the PSF.  We have
demonstrated only that a measurement performed with a circular sinc
aperture will have lower variance than one performed with a `top hat'
aperture.  By considering the PSF, we can determine the signal to
noise ratio as a function of aperture radius.  This is shown for a
double Gaussian with $\sigma_{\mathrm{core}}$=1.5 pixels in the lower
panel of Figure~\ref{fig:nea}.  Curves for both sinc and top hat
apertures are shown.  The peak S/N at
$\sim1.6\times\sigma_{\mathrm{core}} \approx 2.4$ pixels is a known
result for a Gaussian PSF \citep[see for example][]{pritchet81}.
However, as the S/N deteriorates to zero for the top hat aperture, it
levels off to a constant for the sinc aperture.  Thus, though S/N is
not optimal at small radii, it approaches a constant rather
than zero.

\section{Fast Computation of Coefficients}
\label{sec:fft}

The $w_{ij}$ double integral in equation~\ref{eq:intsinc} can be
pre-computed, but is still computationally intensive.  However, by
appealing to Fourier methods a fast computation algorithm is
available.  By expressing the aperture as a top hat function with
radius $\rho$, $\Pi((x^2+y^2)^{1/2}/2\rho)$, and noting that the
$\sinc$ functions are even (i.e., $\sinc(\pi(x-x_i)) = \sinc(\pi(x_i -
x))$), the integral can be expressed as the convolution of a
two-dimensional Cartesian $\sinc$ with a circular top hat, sampled
discretely at $x_i, y_j$:

\begin{equation}
\label{eq:intsincconv}
w_{ij} = \iint\limits \ \Pi(x,y) \ \frac{\sin(\pi (x_i - x))}{\pi (x_i
  - x)}\ \frac{\sin(\pi (y_j - y))}{\pi (y_j - y)}\ \dif x\ \dif y.
\end{equation}

The convolution can be computed quickly as a product of the
Fourier-transformed functions in the wave-number domain.  Both
functions have analytic transforms:

\begin{equation}
\label{eq:tophatpair}
\Pi\left(\frac{(x^2+y^2)^{\frac{1}{2}}}{2\rho}\right) \Leftrightarrow
\frac{ \rho\ \mathrm{J}_1\left( 2\pi\rho(k_x^2+k_y^2)^{\frac{1}{2}}
  \right) } { (k_x^2+k_y^2)^{\frac{1}{2}} },
\end{equation}

\noindent and

\begin{equation}
\label{eq:sincpair}
\sinc\left(\pi(x_i-x)\right)\sinc\left(\pi(y_i-y)\right)
\Leftrightarrow
\Pi\left(\frac{k_x}{\pi}\right) \Pi\left(\frac{k_y}{\pi}\right).
\end{equation}

The former yields an Airy function, and the latter yields a
two-dimensional boxcar that extends to the Nyquist wave numbers.  The
coefficients can be obtained by computing the product of the
wave-number domain components from equations~\ref{eq:tophatpair} and
\ref{eq:sincpair}, and taking the inverse Fourier transform:

\begin{equation}
\label{eq:intsincifft}
w = \mathcal{F}^{-1} \left\lbrace \Pi\left(\frac{k_x}{\pi}\right)
\Pi\left(\frac{k_y}{\pi}\right) \frac{ \rho\ \mathrm{J}_1\left(
  2\pi\rho(k_x^2+k_y^2)^{\frac{1}{2}} \right) } {
  (k_x^2+k_y^2)^{\frac{1}{2}} }.  \right\rbrace
\end{equation}

This can be computed with a fast Fourier transform (FFT).  The array
of values in the wave-number domain extends out to the Nyquist wave
number -- precisely where the boxcar truncates the Airy function.
Thus, instead of performing the real-space integral in
equation~\ref{eq:intsinc}, the $w_{ij}$ coefficients can be found
directly by computing the values for the Airy function
(equation~\ref{eq:tophatpair}), and taking the inverse FFT.  The
transformed $w_{ij}$ coefficients are entirely real-valued, and the
coefficients are mathematically identical to those shown in
Figure~\ref{fig:wijcirc}.

Benchmark calculations of the $w_{ij}$ coefficients for apertures with
radii $r=1-10$ pixels found $\sim 10000\times$ speed increases
($\lesssim1$ ms versus several seconds) when calculations were
performed in the wave-number domain.

\subsection{Shifting the Aperture Centroid}
\label{ssec:shift}

The centroid of a PSF is typically measured with sub-pixel accuracy,
and it is essential that it be possible to recenter our aperture on
non-integer pixel coordinates.  If the $w_{ij}$ coefficients are
pre-computed for integer pixel coordinates, $\sinc$-interpolation can
be used to shift them by the necessary sub-pixel offset,
$\delta_x,\delta_y$.  

As previously discussed, $\sinc$-interpolation is valid only for
functions that are band limited.  We demonstrated that a PSF can be
treated as a band-limited function, but have not done so for the
$w_{ij}$ components.  However, the $w_{ij}$ components represent the
aperture convolved with a $\sinc$ function.  As a $\sinc$ function is
itself band limited, it follows by the convolution theorem, that any
convolution with it will also form a band-limited function.  The
$w_{ij}$ coefficients can safely be shifted by $\sinc$-interpolation.

It is also possible to apply a sub-pixel shift directly to $w_{ij}$
coefficients when they are created with the FFT method.  To perform
aperture photometry on several sources, shifting pre-computed $w_{ij}$
coefficients is faster than directly computing them for each source
individually, but only slightly.  If this approach is taken, shifting
can be achieved easily in the wave-number domain by applying the Shift
Theorem ($f(x-\delta) \Leftrightarrow e^{-i2\pi\delta k}F(k)$).
Dropping the redundant $\Pi(k/\pi)$ terms,
equation~\ref{eq:intsincifft} then becomes

\begin{equation}
\label{eq:wijshifted}
w = \mathcal{F}^{-1} \left\lbrace e^{-i2\pi(\delta_x k_x + \delta_y
  k_y)}\ \frac{ \rho\ \mathrm{J}_1\left(
  2\pi\rho(k_x^2+k_y^2)^{\frac{1}{2}} \right) } {
  (k_x^2+k_y^2)^{\frac{1}{2}} } \right\rbrace.
\end{equation}

Due to the shift, the wave-number domain components are neither
real-valued nor even.  However, the application of the shift term,
$e^{-i2\pi\delta k}$, preserves conjugate symmetry (i.e., if $f(x)$ is
real-valued, then $F(k) = F^{*}(-k)$, where `$^*$' denotes complex
conjugation).  With conjugate symmetry, the spatial domain values
(i.e., in pixel space) are guaranteed to be real-valued.  This is not
surprising given the physical interpretation of a shift in the spatial
domain.

\section{Other Apertures}
\label{sec:otherapertures}

Photometry with any aperture can be performed with this method; the
circular aperture is conventional in observational astronomy, and
conveniently has an analytic form in the wave-number domain.  If the
Fourier transform of the desired aperture can be accurately computed,
it can be substituted for the Airy function in
equations~\ref{eq:intsincifft} and~\ref{eq:wijshifted}.

If the desired aperture has no analytic transform, the $\sinc$
convolution integral in equation~\ref{eq:intsinc} can be computed
directly.  As this is more computationally intensive, it could be
impractical to recompute for each source being measured, and the
integral should be pre-computed and shifted to the pixel coordinates
of the object being measured.

\subsection{Annular Apertures}
\label{ssec:annular}

Due to the addition theorem ($f(x) + g(x) \Leftrightarrow F(k) + G(k)$),
an annular aperture can be constructed trivially by taking a
difference of Airy terms representing the inner and outer radii of the
annulus.  The $w_{ij}$ coefficients for a range of annular apertures
are shown in Figure~\ref{fig:wijannul}.

\middfig{\figWijannul}

\subsection{Elliptical Apertures}
\label{ssec:elliptical}

The special case of an elliptical aperture can be handled by applying
the Similarity Theorem ($f(ax) \Leftrightarrow |a|^{-1}F(k/a)$) to the
circular case.  The compression of a circle along a given axis forms
an ellipse, and the corresponding Fourier transform can be formed by
dilating that of a circle along the appropriate axis.

The Fourier transform preserves rotation; if $f(x,y)$ is rotated
by $\theta$ degrees, $F(k_x,k_y)$ will be rotated by a corresponding
amount in the same sense.  We therefore compress the Airy function
along the $k_y$ axis, and rotate it by the desired amount with a
rotation matrix applied to the $k_x,k_y$ wave numbers.

To demonstrate all the properties presented, $w_{ij}$ coefficients
for a selection of annular elliptical apertures are shown in
Figure~\ref{fig:wijellip}.

\middfig{\figWijellip}

In Appendix~\ref{app:code}, we include a sample of Python code that
demonstrates the method by computing{\highlight growth curves for model
PSFs}.  The coefficient images in Figures~\ref{fig:wijcirc},
\ref{fig:wijannul}, and~\ref{fig:wijellip} were constructed with the
{\tt wijCoefficients()} routine included in the sample code.

\section{Performance and Effectiveness of the Algorithm}
\label{sec:performance}

The accuracy of the method was tested by constructing and measuring
images of sources with known PSFs and fluxes.  Two different tests
were performed.  In the first{\highlight test}, aperture magnitudes
were measured for progressively larger apertures to determine the
growth curves for model PSFs, and these were compared to those of the
known, analytic solutions.{\highlight In addition to the double Gaussian PSF
  used throughout this work, the popular Moffat PSF \citep{moffat69}
  was also tested}.  In the second test, our aperture magnitudes were
compared to values produced by the SExtractor software package
\citep{bertin96}.

\subsection{Measuring the Growth Curves for Double Gaussian and Moffat PSFs}
\label{ssec:growthcurve}

{\highlight For simplicity, analytic double Gaussian and Moffat PSFs were used
  for testing.  Neither is a band-limited function, and different
  widths ($\sigma_{\mathrm{core}}$ values for the double Gaussians and
  FWHM for the Moffat PSFs) were tested to evaluate the behaviour near
  the critical sampling limit, $\sigma_{\mathrm{core}}=1$ pixel.  The
  pixel values for the double Gaussian were computed at coordinates
  $x,y$ with

\begin{equation}
\label{eq:pixelvalues}
p_{ij} =
\frac{1}{2\pi\left(\sigma^2_{\mathrm{core}}+b\sigma^2_{\mathrm{wing}}\right)}
\left[e^{\frac{-(x^2+y^2)}{2\sigma_{\mathrm{core}}^2}} +
  b\ e^{\frac{-(x^2+y^2)}{2\sigma_{\mathrm{wing}}^2}}\right].
\end{equation}

\noindent Here, $\sigma_{\mathrm{wing}} = 2\sigma_{\mathrm{core}}$, and $b=0.1$.
Values for the Moffat PSF were computed with

\begin{equation}
  \label{eq:pixelvaluesMoffat}
  p_{ij} = \frac{\beta-1}{\pi\alpha^2}\left[1.0 + \frac{x^2 +
      y^2}{\alpha^2}\right]^{-\beta}
\end{equation}

\noindent where $\beta=4.765$ \citep{trujillo01c}, and $\alpha =
\mathrm{FWHM}/2(2^{1/\beta}-1)^{1/2}$.  The FWHM used to compute the
Moffat PSF was $2.4670\sigma_{\mathrm{core}}$.  Note that this differs
  slightly from the value of FWHM=$2.3548\sigma$ for a single
  Gaussian.

}

Note that these are a sampling of the PSFs as we do not integrate over
the pixels.  The PSF is considered to be a combination of multiple
contributing components, including the aperture response (an Airy
function), the atmospheric blurring, and the pixel response.

Three widths were evaluated: $\sigma_{\mathrm{core}}$=0.8, 1.0, and
1.2 pixels, corresponding to under-, critical-\footnote{\highlight
  Strictly speaking, there is no `critical' sampling limit for either
  a Gaussian or a Moffat PSF as they are not band-limited functions.
  We use the term loosely, but for all practical intents and purposes
  (as we show) $\sigma_{\mathrm{core}}$=1.2 pixels is quite
  reasonable.}, and well-sampled PSFs.  Sources were planted in the
center of a 16$\times$16 pixel image.  Additional sub-pixel offsets
were applied to test the effects of shifting the aperture.

Sinc-integrated aperture fluxes were measured from the images with
aperture radii, $0 < r < 4$ pixels (0.1 pixel increments).  The known
PSFs (equation~\ref{eq:pixelvalues}) were then integrated directly
over the same apertures to obtain the true theoretical fluxes.  The
structure of the residuals varies depending on the sub-pixel shift
applied.{\highlight For both PSFs}, the smallest residual was
observed when no sub-pixel offset was applied, and the largest
residual was observed when half-pixel offsets were applied to both $x$
and $y$.  Discrepancies are highest for apertures with radii $<1$
pixel.  Growth curves of the measured fluxes are presented with
residuals in Figure~\ref{fig:growthcurves}.  To illustrate the effects
of the sub-pixel shift, the curves shown are those representing the
extrema.

The amplitudes of the residual curves decrease rapidly as sampling
improves.  Discrepancies as high as a few percent can be seen for the
undersampled PSFs, and the method is clearly unsuitable under such
conditions.{\highlight The well-sampled PSF ($\sigma_{\mathrm{core}}
  = 1.2$, FWHM$=3.0$) performs extremely well at all aperture sizes
  (fractional residual less than $0.001$ and $0.002$ for double
  Gaussian and Moffat, respectively)}.

\middfig{\figGrowthcurve}

\subsection{Comparison to SExtractor Aperture Magnitudes}
\label{ssec:sextractor}

We also compared sinc magnitudes to the values produced by the
SExtractor software package \citep{bertin96} version 2.3.2.
SExtractor is one of the most widely used photometry packages in the
astronomy community and is the most suitable algorithm for such a
comparison.

Double Gaussian PSFs were added to the center of $24\times24$ pixel
images.  Tests were performed with well-sampled
($\sigma_{\mathrm{core}}$=1.5 pixel) and under-sampled
($\sigma_{\mathrm{core}}$=0.7 pixel) double Gaussians.  As with
previous tests, the wing component was 2$\times$ wider and{\highlight
  had an amplitude which was 10$\%$ of the core component's
  amplitude}.  Each test was run with and without noise for aperture
radii of $r$=1.0, 2.0, and 5.0 pixels.  The tests performed with noise
included a 100 count sky level, and the pixel values were replaced
with random Poisson variates.{\highlight For simplicity, a gain of 1.0
  was used.}

For a given test, 500 trials were run for fluxes covering several
magnitudes.  In each case, the double Gaussian was given a random
sub-pixel offset.{\highlight The test image was written to a FITS
  file and was measured with both SExtractor and the sinc algorithm.
  In order to ensure that only the photometric algorithms were
  compared, the sinc algorithm was run using the $x,y$ pixel centroids
  as measured by SExtractor.}  The PHOT\_APERTURES parameter was set
to 2$\times$ the radius (SExtractor specifies aperture diameter), and
MAG\_ZEROPOINT was set to an arbitrary value of 25.  For tests
including noise and sky, the known sky level of 100 counts was
subtracted before processing, and SExtractor's BACK\_TYPE parameter
was set to MANUAL to disable internal background estimation.  Results
of the testing are shown in Figures~\ref{fig:sexwellsampled}
(well-sampled PSF), and~\ref{fig:sexundersampled} (under-sampled PSF).

\middfig{\figSexWellSampled}

\middfig{\figSexUnderSampled}

For well-sampled data measured with a large ($r\gtrsim$5 pixel)
aperture, the performance of the two algorithms is similar.  For
smaller apertures, the sinc algorithm outperforms SExtractor.  In the
case of under-sampled data with a large aperture, SExtractor
outperforms.  However, this performance deteriorates for smaller
apertures while the sinc algorithm's does not.  For $r\lesssim$1.0
pixel apertures, the sinc method yields the better result.

\section{Discussion}
\label{sec:discussion}

The method we have described relies on very well-established
mathematical theorems, namely the sampling theorem, and various
theorems associated with the Fourier transform.  However, various
aspects of our analysis require some discussion.

{\highlight Our decision to use pure sinc interpolation in place of a
  Lanczos kernel (a tapered sinc function, see \citet{lanczos56})
  requires some justification.  When we integrate the sinc function
  over the aperture, it is defined within the entire region of
  integration, regardless of the location of the sinc peak (i.e. the
  pixel in $w_{ij}$ being considered).  The value obtained for each
  $w_{ij}$ pixel is therefore more accurate than that which would be
  obtained by a Lanczos, which tapers to zero beyond its region of
  support (typically 4 to 7 pixels).  However, this difference is
  truly negligible.  Our principal reason for using a pure sinc was
  that it offers a convenient Fourier transform (unity throughout our
  region of interest in the wave-number domain).  In order to minimize
  demand on computing resources for e.g. a Kron flux, computation in
  the wave-number domain is essential.  The Fourier transform of a
  Lanczos kernel is sufficiently complicated that the pure sinc is the
  better choice.  }

As there are a variety of PSF models available, an alternative to
measuring aperture fluxes directly would be to fit an appropriate
model and integrate the model over the desired aperture.  When the
object being measured is not a point source, this approach is often
preferred.  In such cases, the objects have different sizes and
shapes, and use of a constant aperture radius would not allow objects
to be meaningfully compared.  Model fluxes can frequently be the best
choice in such situations.  However, for the purpose of calibration,
the use of a model introduces an unnecessary source of potential
systematic error.  A poorly chosen model will never fit properly and
will corrupt the aperture correction and thus the zero point of the
calibration.

As a more practical verification of the method, the sinc-interpolated
aperture fluxes are being used to compute aperture fluxes, and to
determine the aperture correction for the LSST photometric pipeline
that is currently under development.  The sinc method has been found
to perform extremely well, providing
$\left<m_{\mathrm{PSF}}-m_{\mathrm{ap}}\right> \lesssim 0.001$
(verifying aperture correction), and
$\left<m_{\mathrm{ap}}-m_{\mathrm{cat}}\right> \lesssim 0.001$
(verifying measurement accuracy).  Here, $m_{\mathrm{PSF}}$ and
$m_{\mathrm{ap}}$ are PSF and (sinc) aperture magnitudes, and
$m_{\mathrm{cat}}$ is a catalog magnitude corresponding to the known
input flux in a simulated image.  Several hundred simulated images
have been generated, processed, and tested, covering a variety of
observing conditions (cloud, seeing, etc).{\highlight The LSST
  simulator uses a ray-tracing algorithm to trace photons through a
  simulated atmosphere and through simulated optics to produce each
  test image \citep[][in preparation]{peterson13}.  The ray-tracing
  simulations include realistic field-dependent and
  wavelength-dependent effects that result in complex point-spread
  functions and photometric variation that challenge photometry
  algorithms.  }.  The aperture magnitudes show no systematic trends
with magnitude, and the variability in shape and PSF width observed
across a focal plane of individual CCD images is handled well by a 2D
polynomial aperture correction.{\highlight An example of sinc aperture
  photometry performed on LSST simulated data is shown in
  Figure~\ref{fig:lsst}. } As yet, no failure modes have been observed
for the algorithm.

\middfig{\figLsst}

As stated earlier, any error introduced as a result of a
non-band-limited PSF will not be systematic.  Even if a constant
radius aperture is used for aperture photometry, the error will depend
on the sub-pixel shift of the PSF and will thus be different for each
measured source.  Adequate sampling is essential, but we believe our
tested limit of $\sigma > 1.2$ pixels{\highlight (FWHM $> 3.0$
  pixels)} is a reliable benchmark.

\section{Summary}
\label{sec:summary}

We have developed an algorithm for performing aperture photometry on
critically sampled astrophysical images.  The flux is numerically
integrated over the desired aperture, with sinc-interpolation used to
reconstruct values between pixels.  As the pixel values can be
factored out of the computationally intensive integral, only the
$\sinc$ functions need to be integrated, and these can be pre-computed
and applied to an arbitrary number of aperture flux measurements.

The aperture flux is ultimately computed as a weighted sum of pixel
values, with the sinc-integrals providing the weights.  The weight
values are non-zero outside the nominal cut-off radius of the
aperture.  We find that {\bfseries an aperture flux must be measured
  in a footprint with a border extending $\sim$4 pixels beyond the
  aperture limits to ensure accuracy}.

Using Fourier methods, we have shown that the integral of the
sinc-functions over the aperture is a convolution integral, and can be
computed quickly as a product in the wave-number domain.  

As the method relies on sinc-interpolation based on the
sampling-theorem, a band-limited PSF is an implicit assumption.  A
rule of thumb for this requirement to be satisfied is that {\bfseries
  a double-Gaussian PSF must be sampled such that $\sigma > 1.2$
  pixels, or equivalently{\highlight FWHM $> 3.0$ pixels}}. No
Gaussian-based function is truly band limited, but this rule of thumb
reduces the photometric error contributed by the aperture to $< 0.001$
mag.  If the requirement is not satisfied, the resulting error depends
on the sub-pixel position of the PSF and cannot be corrected during
calibration.  The error is also a function of aperture radius, and the
band-limited requirement must be satisfied to produce an accurate
growth curve.

Comparison with SExtractor demonstrated that the sinc method provides
equal performance for large apertures ($r\gtrsim$5 pixel), and much
better performance for smaller apertures.  However, for under-sampled
data, the best results were obtained with SExtractor using a large
aperture.

The method performs equally well for annular and elliptical apertures,
and can be adapted for any aperture.  To demonstrate the method, a
short example has been coded in Python and is presented in
Appendix~\ref{app:code}.

In closing, we note that the Sloan Digital Sky Survey's photometric
pipeline used a version of this technique for the measurement of
aperture fluxes.  In that version, the $w_{ij}$ components (see
equation~\ref{eq:intsincconv}) were pre-computed as real-space 2D
integrals prior to frame processing, and were Lanczos-shifted to be
recentred on each source for flux measurement.  As the SDSS
photometric catalog is among the most widely used data sets in the
astronomy community, the effectiveness of the algorithm has been very
well demonstrated in practice.  The method is now being used in the
LSST photometric pipeline under development, with the only difference
being the use of Fourier methods to more efficiently generate the
$w_{ij}$ components.  Tests on simulated LSST images have shown the
algorithm to perform efficiently (i.e., quickly) and produce
Poisson-limited accuracy.

\newcommand{\ack}{ 

We wish to thank our LSST colleagues Pat Burchat, Seth Digel, Jon
Thaler, Phil Marshall, and Rachel Mandelbaum for their comments and
advice; and we wish to thank Emmanuel Bertin for his many valuable
suggestions.
  
Support for this research was provided by both the SDSS and LSST
projects.
  
Funding for the SDSS and SDSS-II has been provided by the Alfred
P. Sloan Foundation, the Participating Institutions, the National
Science Foundation, the U.S. Department of Energy, the National
Aeronautics and Space Administration, the Japanese Monbukagakusho, the
Max Planck Society, and the Higher Education Funding Council for
England. The SDSS Web Site is http://www.sdss.org/.

The SDSS is managed by the Astrophysical Research Consortium for the
Participating Institutions. The Participating Institutions are the
American Museum of Natural History, Astrophysical Institute Potsdam,
University of Basel, University of Cambridge, Case Western Reserve
University, University of Chicago, Drexel University, Fermilab, the
Institute for Advanced Study, the Japan Participation Group, Johns
Hopkins University, the Joint Institute for Nuclear Astrophysics, the
Kavli Institute for Particle Astrophysics and Cosmology, the Korean
Scientist Group, the Chinese Academy of Sciences (LAMOST), Los Alamos
National Laboratory, the Max-Planck-Institute for Astronomy (MPIA),
the Max-Planck-Institute for Astrophysics (MPA), New Mexico State
University, Ohio State University, University of Pittsburgh,
University of Portsmouth, Princeton University, the United States
Naval Observatory, and the University of Washington.

LSST project activities are supported in part by the National Science
Foundation through Governing Cooperative Agreement 0809409 managed by
the Association of Universities for Research in Astronomy (AURA), and
the Department of Energy under contract DE-AC02-76-SFO0515 with the
SLAC National Accelerator Laboratory. Additional LSST funding comes
from private donations, grants to universities, and in-kind support
from LSSTC Institutional Members.  

}

\ifthenelse{\boolean{mn2e}}{
\section*{Acknowledgments}
\ack
}{
  \acknowledgments \ack
}

\appendix

\section{The Fractional Aliased Power in a Sampled Gaussian}
\label{app:aliasedpower}

{\highlight The principal requirement for the sinc reconstruction to be valid is
that the PSF be band-limited.  To test this assumption analytically,
we approximate the PSF, $p(r)$, as a circular bivariate Gaussian,
which has a convenient analytic Hankel transform $P(q)$ (circularly
symmetric form of the Fourier transform):

\begin{equation}
\label{eq:gaussPsf}
p(r) = e^{-r^2/2\sigma^2}  \Leftrightarrow  P(q) = \sigma^2 e^{-q^2 \sigma^2/2},
\end{equation}

\noindent where $\Leftrightarrow$ denotes a transform pair.  

The aliased power beyond the band-limit for a Gaussian ($>\pi$ when
$\sigma$ is in pixel units) is

\begin{equation}
\label{eq:intPower}
\mathrm{aliased\ power} = 2\pi \int_\pi^\infty \sigma^4 e^{-q^2 \sigma^2}\ q\ \dif q = \pi \sigma^2 e^{-\pi^2 \sigma^2},
\end{equation}

\noindent where the integrand is the power from
equation~\ref{eq:gaussPsf}.  Under the assumption that the pixels are
square, the convenient circular symmetry used here is an
approximation, but a very good one.  The Fourier transform of a real
PSF should be integrated in two dimensions with limits $k_{x,y} <
-\pi, k_{x,y} > \pi$, so our use of circular symmetry ($q =
(k_x^2+k_y^2)^{1/2} > \pi$) is conservative in that it omits power
that is not actually lost in Cartesian coordinates (i.e., the corners
of the region in the wave-number domain where $\pi < (k_x^2 +
k_y^2)^{1/2}$).

Next, we translate the aliased power to a corresponding error in flux
using Rayleigh's Theorem\footnote{Rayleigh's Theorem is the continuous
  form of the (often more familiar) discrete summation, Parseval's
  Theorem.}  applied to the Hankel transform:

\begin{equation}
\label{eq:parseval}
\int_{0}^{\infty} |p(r)|^2\ r\ \dif r = \int_{0}^{\infty} |P(q)|^2\ q\ \dif q.
\end{equation}

The aliased power equals the integral of the squared error
$|\epsilon(r)|^2$, which that power represents in the spatial domain.
By taking the square root, we obtain a flux error
estimate \footnote{We are interested in the total flux error, and we
  do not divide by the aperture area to obtain a per-pixel
  root-{\itshape mean}-squared (RMS) value.  The error shown is a
  root-{\itshape sum}-squared value.}:

\begin{eqnarray}
\label{eq:parseval}
\mathrm{error} &=& \left(2\pi \int_{0}^{\infty} |\epsilon(r)|^2\ r\ \dif r\right)^{1/2}\\
 &=& \left(2\pi \int_{\pi}^{\infty} \sigma^4 e^{-q^2\sigma^2} q\ \dif q\right)^{1/2}\\
 &=& \pi^{1/2} \sigma e^{-\pi^2\sigma^2/2}.
\end{eqnarray}

The total integrated flux is $2\pi\sigma^2$, and the error shown as a fraction is

\begin{equation}
\frac{\mathrm{error}}{\mathrm{total\ flux}} = \frac{e^{-\pi^2\sigma^2/2}}{2\pi^{1/2}\sigma}.
\label{eq:appfluxerr}
\end{equation}

Equation~\ref{eq:appfluxerr} can be used as an estimate of the systematic
flux error introduced by aliased power.  Three assumptions were
implicitly made to simplify the estimate: (1) circular symmetry can be
used as an approximation in equation~\ref{eq:intPower}, (2) the power
above the Nyquist wave number is simply lost rather than being aliased
back into the pass band, and (3) the flux being integrated is the
total flux integrated to infinity rather than that within a limited
aperture radius.  All of these assumptions tend to {\itshape
  over}estimate the error, and equation~\ref{eq:appfluxerr} can be
regarded as providing an upper limit on the systematic flux error.

} 

\section{Sample Code to Demonstrate Sinc Aperture Photometry}
\label{app:code}

The following sample of Python code demonstrates the sinc method for
performing aperture photometry.  Python is an open-source scripting
language, and is readily available on most computing platforms (see
{\tt http://www.python.org}).  Our example uses the {\tt numpy} and
{\tt scipy} modules (available at {\tt http://numpy.scipy.org/} and
{\tt http://www.scipy.org/}, respectively).

Due to the nature of Python, the indentation structure is essential.
The authors will gladly provide a digital copy upon request.  Note
that the code is intended only as an example, and is not written with
efficiency in mind.  Computing performance can be greatly improved by
transcribing the code to C, C++, or some other compiled language.  A
C++ version of the algorithm is included in the LSST code stack.  For
information on downloading the LSST source code see {\tt
  www.lsst.org}).

The following command will run the example (assuming the script has
been named {\tt sinc\_phot.py}), which will plant a Gaussian PSF with
$\sigma=1.2$ centred at $x,y=16,16$ in a $n=32\times 32$ pixel image,
and will compute a growth curve for it with the sinc-aperture method.
Output is printed to the screen (i.e., {\tt stdout}).  \\

{\tt
\noindent ./sinc\_phot.py 32 16.0 16.0 1.2\\
(i.e., ./sinc\_phot.py n x y $\sigma$)\\
}

The function {\tt wijCoefficients()} is capable of computing annular
and/or elliptical aperture coefficients, but values for the
ellipticity, position angle ({\tt theta}), and inner radius are set to
0.0 in the example.

{\scriptsize 
\begin{verbatim}
#!/usr/bin/env python
import sys, numpy, scipy.special as special

def gaussianImage(n, center_x, center_y, sigma):
    image = numpy.zeros([n, n], dtype=float)
    for ix in range(n):
        for iy in range(n):
            x, y = ix - center_x, iy - center_y
            A = 1.0/(2.0*numpy.pi*sigma**2)
            image[iy,ix] = A*numpy.exp(-(x**2 + y**2)/(2.0*sigma**2))
    return image

def wijCoefficients(n, x, y, rad1, rad2, theta, ellipticity):
    wid, xcen, ycen = n, (n+1)/2, (n+1)/2
    dx, dy = x - xcen, y - ycen
    if n%2:
        dx, dy = dx + 1, dy + 1
        
    ftWij = numpy.zeros([wid, wid], dtype=complex)
    scale     = 1.0 - ellipticity
    for iy in range(wid):
        ky = float(iy - ycen)/wid
        for ix in range(wid):
            kx = float(ix - xcen)/wid
            
            # rotate and rescale
            cosT, sinT = numpy.cos(theta), numpy.sin(theta)
            kxr, kyr =  kx*cosT + ky*sinT, scale*(-kx*sinT + ky*cosT)
            k = numpy.sqrt(kxr**2 + kyr**2)
            
            # compute the airy terms, and apply shift theorem
            if k != 0.0:
                airy1 = rad1*special.j1(2.0*numpy.pi*rad1*k)/k
                airy2 = rad2*special.j1(2.0*numpy.pi*rad2*k)/k
            else:
                airy1, airy2 = numpy.pi*rad1**2, numpy.pi*rad2**2
            airy = airy2 - airy1
            phase = numpy.exp(-1.0j*2.0*numpy.pi*(dx*kxr + dy*kyr))
            
            ftWij[iy,ix] = phase*scale*airy

    ftWijShift = numpy.fft.fftshift(ftWij)
    wijShift   = numpy.fft.ifft2(ftWijShift)
    wij        = numpy.fft.fftshift(wijShift)
    return wij.real

if __name__ == '__main__':
    n = int(sys.argv[1])
    x, y, psf_sigma = map(float, sys.argv[2:])
    radius_inner, theta, ellipticity = 0.0, numpy.pi*(0.0)/180.0, 0.0
    
    psf = gaussianImage(n, x, y, psf_sigma)
    for radius in numpy.arange(radius_inner + 0.1, 6.0*psf_sigma, 0.1):
        wij = wijCoefficients(n, x, y, radius_inner, radius, theta, ellipticity)
        flux_measured = (psf*wij).sum()
        flux_analytic = (1.0 - numpy.exp(-radius**2/(2.0*psf_sigma**2)))
        print radius, flux_measured, flux_analytic
\end{verbatim}
}

\bibliographystyle{mn2e}
\bibliography{mnras-journal,bickerton}

\tailfig{\figSamplingtheorem}
\tailfig{\figFluxerr}
\tailfig{\figWijcirc}
\tailfig{\figNea}
\tailfig{\figFootprint}
\tailfig{\figWijannul}
\tailfig{\figWijellip}
\tailfig{\figGrowthcurve}
\tailfig{\figSexWellSampled}
\tailfig{\figSexUnderSampled}
\tailfig{\figLsst}

\label{lastpage}

\end{document}